\documentclass[sigconf]{acmart}

\usepackage{xcolor}
\usepackage{todonotes}
\usepackage[framemethod=TikZ]{mdframed}
\usepackage{tabularray}
\newmdenv[
  backgroundcolor=gray!10,
  linecolor=gray,
  linewidth=1pt,
  roundcorner=5pt,
  skipabove=12pt,
  skipbelow=12pt,
  innerleftmargin=10pt,
  innerrightmargin=10pt,
  innertopmargin=10pt,
  innerbottommargin=10pt
]{mybox}
\usepackage{caption}

\AtBeginDocument{%
  }

\setcopyright{acmlicensed}
\copyrightyear{2025}
\acmYear{2025}
\acmDOI{XXXXXXX.XXXXXXX}
\acmConference[SIGIR]{SIGIR 2025 LiveRAG Challenge}{July 17, 2025}{Padua, IT}
\acmISBN{978-1-4503-XXXX-X/2018/06}

\begin{document}

\title{RAGtifier: Evaluating RAG Generation Approaches of State-of-the-Art RAG Systems for the SIGIR LiveRAG Competition}


\author{Tim Cofala}
\affiliation{%
  \institution{L3S Research Center}
  \city{Hannover}
  \country{Germany}}
\email{tim.cofala@l3s.de}

\author{Oleh Astappiev}
\affiliation{%
  \institution{L3S Research Center}
  \city{Hannover}
  \country{Germany}}
\email{astappiev@l3s.de}

\author{William Xion}
\affiliation{%
  \institution{L3S Research Center}
  \city{Hannover}
  \country{Germany}}
\email{william.xion@l3s.de}

\author{Hailay Teklehaymanot}
\affiliation{%
  \institution{L3S Research Center}
  \city{Hannover}
  \country{Germany}}
\email{teklehaymanot@l3s.de}

\renewcommand{\shortauthors}{Trovato et al.}


\begin{abstract}

Retrieval-Augmented Generation (RAG) enriches Large Language Models (LLMs) by combining their internal, parametric knowledge with external, non-parametric sources, with the goal of improving factual correctness and minimizing hallucinations.
The LiveRAG 2025 challenge explores RAG solutions to maximize accuracy on DataMorgana's QA pairs, which are composed of single-hop and multi-hop questions. The challenge provides access to sparse OpenSearch and dense Pinecone indices of the Fineweb 10BT dataset. It restricts model use to LLMs with up to 10B parameters and final answer generation with Falcon-3-10B.
A judge-LLM assesses the submitted answers along with human evaluators. By exploring distinct retriever combinations and RAG solutions under the challenge conditions, our final solution emerged using \mbox{InstructRAG} in combination with a Pinecone retriever and a BGE reranker. Our solution achieved a correctness score of $1.13$ and a faithfulness score of $0.55$ in the non-human evaluation, placing it overall in third place in the SIGIR 2025 LiveRAG Challenge \cite{carmel2025sigir2025liverag}.
The RAGtifier code is publicly available\footnote{https://git.l3s.uni-hannover.de/liverag/ragtifier}.

\end{abstract}

\keywords{Retrieval Augmented Generation, Adaptive Retrieval, Self-Reflection, LLMs-as-Judges}



\maketitle

\section{INTRODUCTION}\label{sec:introduction}

Retrieval Augmented Generation (RAG) has emerged as a key technique for enhancing LLM performance in question answering (QA) by incorporating external knowledge \cite{zhang2024retrievalqaa, mallen2023trustlanguagemodelsinvestigating}. The LLM prompt is enriched with retrieved information to mitigate issues related to unknown or sparse knowledge within the model itself \cite{press2023measuringnarrowingcompositionalitygap}.

On the live challenge day of the \mbox{LiveRAG} Challenge, participants are provided with 500 DataMorgana-generated \cite{DBLP:journals/corr/abs-2501-12789} questions. The generation of answers with Falcon-3-10B \cite{tiiFalcon} and their submission is limited to a two-hour time slot. The organizers of the \mbox{LiveRAG} Challenge also provide access to the sparse search engine OpenSearch \cite{opensearch} and the dense vector database Pinecone \cite{pineconeVectorDatabase}, both populated with data from Fineweb-10BT \cite{huggingfaceHuggingFaceFWfinewebDatasets}. Subsequently, the State-of-the-Art (SotA) LLM, Claude-3.5-Sonnet, evaluates the submitted answers according to correctness~\cite{pradeep2025greatnuggetrecallautomating} and faithfulness~\cite{es2025ragasautomatedevaluationretrieval} scores. The top-ranked submissions are also manually evaluated to determine the final ranking.

\begin{figure}
 \includegraphics[width=\columnwidth]{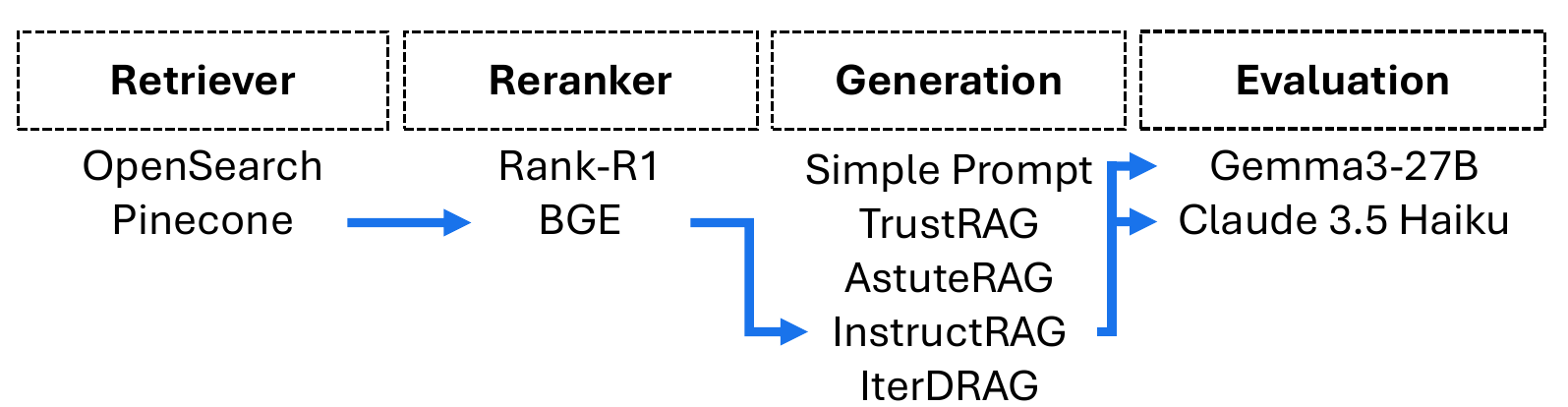}
 \caption{Overview of the RAG solutions and components under consideration. The highlighted path highlights our submitted solution based on the performance evaluation by Gemma-3-27B and Claude-3.5-Haiku.}
 \label{fig:pipeline}
\end{figure}

Our approach (Figure \ref{fig:pipeline}) combines the cross-evaluation of SotA RAG solutions with research insights on the optimization of distinct RAG components. We benchmarked five different generation strategies, two retrievers, two rerankers, and context-ordering techniques using two evaluation LLMs and various single- and multi-hop questions generated by DataMorgana for our internal benchmark.





\section{RELATED-WORK}\label{sec:introduction}
In this section, we outline the relevant components of a complete RAG pipeline and associated concepts.

\paragraph{\textbf{Retriever}}
It collects and evaluates information to expand LLM queries using sparse (lexical) or dense (semantic) methods~\cite{DBLP:conf/sigir/CuconasuTSFCMTS24, DBLP:journals/access/HambardeP23}.
Sparse methods like BM25 \cite{DBLP:journals/ftir/RobertsonZ09} are used in OpenSearch \cite{opensearch}, while dense methods may use ANN-based systems like Pinecone~\cite{pineconeVectorDatabase}. Recent work compares retriever performance on QA tasks \cite{DBLP:conf/sigir/HulyPCKM24} and investigates context length and document order effects \cite{DBLP:conf/sigir/CuconasuTSFCMTS24}.

\paragraph {\textbf{Reranker}}
This component rearranges the retrieved documents to enhance contextual relevance~\cite{DBLP:conf/sigir/ZamaniB24}. Recent research in reranking addresses
faster comparison methods~\cite{DBLP:conf/sigir/ZhuangZKZ24} or improved search relevance~\cite{bge-m3, venktesh2025sunarsemanticuncertaintybased}.

\paragraph{\textbf{Generation}}
The generation process consists of using the retrieved documents as context to generate a coherent and relevant answer to the question. We select five SotA RAG solutions \mbox{(Figure \ref{tab:Gemma3-eval-500-basics})} that exclusively consider retrieval-augmented prompts \cite{DBLP:conf/nips/LewisPPPKGKLYR020} or additionally reflect on retrieved passages \cite{wei2025instructrag, zhou2025trustragenhancingrobustnesstrustworthiness}. Further approaches include the comparison of passages with the parametric knowledge of LLMs \cite{wang2024astuteragovercomingimperfect} or the execution of retrievals and generation in several rounds \cite{DBLP:conf/iclr/YueZB0JZ0WWB25}.






\paragraph{\textbf{Evaluation}}
Judging generated answers by LLMs \cite{DBLP:conf/nips/ZhengC00WZL0LXZ23} became an alternative to simple string matching and inefficient, expensive human evaluators \cite{chiang-lee-2023-large}.
Influencing factors of LLM-judges include bias through prompt styles \cite{DBLP:journals/corr/abs-2409-15268}, answer-length \cite{DBLP:journals/corr/abs-2404-04475} or cross-capability performance \cite{DBLP:journals/corr/abs-2409-19951}.
Mitigation approaches suggest the expensive use of high-performance LLMs \cite{DBLP:journals/corr/abs-2407-18370}, finetuning \cite{DBLP:journals/corr/abs-2310-17631} or confidence estimation \cite{DBLP:journals/corr/abs-2407-18370}.
We select two judge-LLMs for our performance metric to evaluate the generated answers.


\section{RAG COMPONENTS}
\label{sec:rag-structure}
In this section, we provide a detailed overview of the RAG components considered for the \mbox{LiveRAG} Challenge, followed by a description of the components used in our submission.

\subsection{DataMorgana}
DataMorgana \cite{DBLP:journals/corr/abs-2501-12789} is a novel approach to generate QA-pairs from documents by defining diverse user and question categories to create highly customizable synthetic benchmarks.
We define additional user and question categories (Figure \ref{box:datamorgana}) to increase the variety of generated questions, thereby further challenging the answer generation capabilities. From a pool of 10,000 generated questions, we created a dataset of 500 randomly selected QA-pairs, evenly distributed across single- and multi-document subsets.

\subsection{Retriever}
Our RAG solution considers the dense retriever implemented with Pinecone. This choice is based on experiments with the QA-pairs we generated. We compared both provided indices, where the dense retriever demonstrated faster response times and a higher retrieval rate of gold documents (@k), both with and without an additional reranker.

\subsection{Reranker}


We investigate the performance of BGE-M3~\cite{bge-m3} and Rank-R1~\cite{DBLP:journals/corr/abs-2503-06034}. Both rerankers aim to improve document relevance by reordering retrieved documents according to their relevance to the input query, thereby enhancing the quality of the context provided to the generation model.

We investigate the key performance characteristics of the BGE-M3 reranker, focusing on its latency in combination with different amounts of retrieved documents, its ability to handle diverse queries and document lengths for context understanding, and its ranking accuracy with distinct queries and retriever settings to determine the optimal configurations.

In exploring alternative SotA reranking methods, we also investigate Rank-R1 \cite{DBLP:journals/corr/abs-2503-06034}, a novel LLM-based reranker notable for its explicit reasoning capabilities. However, Rank-R1's application was ultimately deemed impractical due to its processing time, which can take up to $100s$ for a single query, making it unsuitable for the time constraints of the \mbox{LiveRAG} Challenge.

\subsection{Generation}

We consider recent advances in RAG and cross-evaluate various answer generation approaches with distinct retriever and reranker settings to compare their performance on DataMorgana-generated QA-pairs. We use a non-finetuned Falcon-3-10B LLM for all generation tasks, with a temperature setting of $0.1$. We consider the generation prompts from the following RAG approaches:

\paragraph{Simple Prompt} The instruction utilizes direct-input augmentation for answer generation, combining the retrieved documents followed by the query~\cite{DBLP:conf/nips/LewisPPPKGKLYR020}.

\paragraph{TrustRAG} The solution proposes a three-step process where the retrieved information is compared against the parametric knowledge to filter out malicious or irrelevant documents, aiming to enhance the security and reliability of answer generation against retrieval-influenced corpus poisoning attacks~\cite{zhou2025trustragenhancingrobustnesstrustworthiness}.

\paragraph{InstructRAG} That strategy introduces a framework for explicit context denoising in retrieval-augmented generation through a two-phase methodology~\cite{wei2025instructrag}.
First, rationale generation utilizes the LLM's instruction-following capabilities to identify relevant information in noisy inputs.
Second, explicit denoising learning employs synthesized rationales as demonstrations or training data that enable effective denoising strategies.

\paragraph{Astute RAG} A framework addressing imperfect retrieval results through a three-phase process: adaptive elicitation of internal model knowledge, source-aware knowledge consolidation, and reliability-based answer finalization~\cite{wang2024astuteragovercomingimperfect}.
In contrast to conventional RAG implementations, Astute RAG explicitly identifies and resolves knowledge conflicts between the model's parametric knowledge and the retrieved information and adaptively combines the most reliable elements from each source.

\paragraph{Iterative Demonstration-Based RAG (IterDRAG)} An iterative approach based on Demonstration-based RAG (DRAG)~\cite{DBLP:conf/nips/BrownMRSKDNSSAA20}, where contextualized examples guide the LLM in its long-context usage~\cite{DBLP:conf/iclr/YueZB0JZ0WWB25}.
IterDRAG extends this approach by incorporating a multi-round question refinement process, specifically targeting multi-hop questions.
It decomposes the main question into sub-queries, generates an answer for each sub-question, which can additionally be retrieved independently, and ultimately constructs the final prompt containing the original question, the complete set of retrieved documents, follow-up questions, and intermediate answers.

All RAG generation approaches, with the exception of IterDRAG due to its inherent complexity, utilize the inverted context ordering proposed by Cuconasu et al.~\cite{cuconasu2024power}. In this ordering, the retrieved or reranked documents are arranged in descending order of relevance, with the highest-ranked document placed immediately before the question.

\subsection{Evaluation}
Gemma-3-27B \cite{googleGemmaOpen} serves as the primary evaluation model in our RAG Challenge pipeline.
Considering the expensive use of the best-performing models, which are also the best-performing judges \cite{DBLP:journals/corr/abs-2410-13341}, we searched for the smallest, yet best-performing LLM on the Chatbot Arena \cite{openlmChatbotArena}. This approach also aligns with the recommendation against using the same LLM for both answer generation and evaluation \cite{DBLP:journals/corr/abs-2410-02736}.
Our selection is therefore based on its competitive Chatbot Arena Elo score of 1341, which ranks it as the smallest among the top open models.
While acknowledging the potential drawbacks associated with this specific model, given the complexities of investigating biases and unexpected weaknesses in LLMs-as-Judges~\cite{DBLP:journals/corr/abs-2410-02736}, we proceeded with Gemma-3-27B. Additionally, we evaluate the candidate systems using Claude-3.5-Haiku~\cite{anthropic2024claude}, as this model family is utilized for the final evaluation in the \mbox{LiveRAG} Challenge.

Furthermore, we consider different proposed evaluation prompting techniques to investigate influencing factors on LLMs-as-Judges evaluations.
The first evaluation prompt (simple comparison) compares only the Falcon-3-10B generated answer with DataMorgana's generated ground-truth answer \cite{DBLP:conf/nips/ZhengC00WZL0LXZ23}.
The prompt directly instructs the LLM to decide which answer is better or if it's a tie after providing a brief explanation of each answer's most important aspects.
The second evaluation prompt is derived from CRAG \cite{DBLP:conf/nips/YangSXSBCCGJJKM24}, which employs several metrics to define a good answer, such as conciseness, correctness, and support from retrieved documents.
\label{sec:live_rag_prompt}
The third evaluation approach extends these methods by employing two distinct metrics, a correctness score and a faithfulness score, with a range of 4 and 3 possible values per metric, respectively.
This evaluation prompt (Figure \ref{box:claude-eval}), hereafter referred to as the \textit{LiveRAG} prompt, combines the evaluation strategy specified by the \mbox{LiveRAG} Challenge \cite{tiiLiveRAGChallenge} with generated and ground truth information.

\section{EXPERIMENTS}\label{sec:experiments}
Now, we discuss the results for the components we considered, starting with insights from the use of DataMorgana and LLMs-as-Judges.
We then move on to a general investigation of retriever performance and retrieval-influenced reranking, and finally describe the results of the cross-evaluation that led to our LiveRAG solution.

\subsection{DataMorgana Question Generation}
The question generation process with DataMorgana yields an overall solid alignment of questions, documents, and answers.
During question generation, we observed that user and question categories are more likely to influence the generated question as desired if the category description emphasizes \textbf{how it should behave} rather than \textbf{what it should represent}.
For instance, a categorization of a \emph{Spy} with a behavioral description like \emph{hides his true intentions by omitting information, giving misleading details, or communicating in encrypted form} yields more expected results compared to representational description like \emph{secretly gathers information, often for a government or organization, typically about enemies or competitors. Spies use covert methods to obtain intelligence}.

\subsection{Evaluating Prompts and Judges}

First, we investigate the impact of three different evaluation prompts. We select a small subset of $100$ DataMorgana-generated QA- pairs to cross-evaluate the influence of these prompts on the overall evaluation metric. We use Falcon-3-10B to generate answers by providing: 1. query only, 2. golden document and query, and 3. OpenSearch@k and query, where \mbox{$k \in \{1,5,10,20,50\}$}. Subsequently, we evaluate all prompts using Gemma-3-27B with the generated answer and any additional necessary prompt information.
At this phase, we use the OpenSearch retriever due to its superior performance on DataMorgana queries with fewer than 50 retrieved documents, which aligns with Falcon-3-10B's limited context window of up to 50 retrieved passages.

In general, the answers generated using the query and gold documents achieved the highest performance across all prompt styles.
In the simple comparison prompt, answers generated with retrieved documents were penalized, as scores decreased as retrieval@k increased. While only specifying the query yielded the second-best performance, surpassed only by the inclusion of the golden document.
The CRAG prompt favors retrieval@k with $k\in\{5,10\}$ over $k\in\{20,50\}$ and penalizes answers generated with only the query.
The LiveRAG prompt (Figure \ref{box:claude-eval}) favors an increase in retrieval@k for the correctness metric. Compared to the simple comparison and the CRAG prompt, query-only answers are scored almost as correctly as the best-performing answer generation strategies, but these answers are penalized with the lowest faithfulness scores among all generation settings.

Since these different prompts provide comparable results with their biased characteristics, we choose the LiveRAG prompt because of its detailed assessment of correctness and faithfulness and its similarity to the LiveRAG Challenge evaluation methodology.

Investigating the performance of Gamma-3-as-a-Judge, we compare a subset of questions judged by Gemma-3-27B against the judgments of Claude-3.5-Haiku. To do this, we selected samples that were rated as poor, fair, and good by Gemma-3-27B (using the LiveRAG prompt) and re-evaluated them with Claude-3.5-Haiku. As a result, the poor and good samples evaluated by Gemma-3-27B yielded nearly identical correctness and faithfulness scores when judged by Claude. For the mediocre samples, we notice a minor shift towards lower scores from Claude.

\subsection{Retriever Performance}
Considering the time constraints of the LiveRAG Challenge and the influence of golden documents in retrieval@k, we investigate the runtime of the retrieval and the number of golden documents returned at distinct retrieval@k values for the provided OpenSearch and Pinecone indices.
Figure \ref{fig:gold_docs_ret} shows that the number of gold documents increases continuously with higher retrieval@k. Notably, Pinecone outperforms OpenSearch at $@k=20$ for multi-hop questions and at $@k=50$ for single-hop questions.
Runtime measurements (Figure~\ref{fig:runtimes}) reveal that Pinecone is faster for all retrieval@k at $k=[1.600]$. OpenSearch takes $0.12s$ and Pinecone $0.15s$ at $@k=1$, scaling nearly linearly up to $k=600$, where OpenSearch takes $0.9s$ and Pinecone $0.58s$. We decide to proceed with Pinecone due to its performance after reaching Falcon-3-10B's context limit of about $50$ retrieved documents. Consequently, we check the performance increase by using a reranker in combination with retriever@$k>50$.

\subsection{Reranker Performance}

Evaluating the impact on the performance of BGE reranker, we measure the runtime affected by retrieval@k and reranker@k and report the percentage of remaining golden documents. We consider retrieval@k for $k\in[1,300]$ and reranker@k for $k\in\{1,3,5,10,20\}$, allowing Falcon-3-10B with its limited context length to be used for all generation tasks. We cross-evaluate various retrieval and reranker settings to find configurations that perform better than retrieval@k alone in terms of the number of retrieved golden documents for single- and multi-hop questions within a reasonable runtime.
\\
The runtime of BGE (Figure \ref{fig:runtimes}) increases with the number of retrieved documents. Further experiments reveal that BGE takes $\sim11.2s$ to rerank 400 retrieved documents. Considering the \mbox{LiveRAG} time constraints and available computational resources, we limit further experiments to 300 retrieved documents, which takes $\sim8.6s$ per reranking operation per question.
If we increase k up to the context limit of $k=50$ (Figure \ref{fig:gold_docs_ret}), the percentage of returned gold documents increases due to the query alone. We hypothesize an increasing RAG performance due to more golden documents are present in the context when a higher retrieval@k is used in combination with a reranker set to $k\leq20$. Therefore, we searched within the subset of viable retriever and reranker settings for a configuration that outperforms retrieval@50 in terms of gold documents@k (Figure \ref{fig:gold_docs_ret_2}).
For single-hop questions using Pinecone@300 with BGE@10 and OpenSearch and Pinecone@[100, 300] with BGE@20, more gold documents remaining in the reranked set compared to using OpenSearch or Pinecone@50 alone.
Similarly, for multi-hop questions, this occurs for OpenSearch and Pinecone@[100, 300] with BGE@20.




\subsection{Final System Performance}

With our insights into retriever and reranker performance, we cross-evaluate various settings. We use Pinecone@$\{100, 200, 300\}$ with BGE@$\{5,8,10,12\}$ combined with optional inverted context order for each RAG solution. Due to the iterative context generation of IterDRAG, we omitted the context ordering and tested Pinecone@$\{100,200\}$ with BGE@$\{5,10\}$ for the initial retrieval step. For additional retrieval steps in IterDRAG, we considered Pinecone@$200$ with BGE@$\{4,5\}$ for a maximum of four and five iterations respectively, and BGE@3 for six iterations.

Considering the LiveRAG constraints, increasing retrieval@k for a fixed rerank@k does not consistently lead to better performance. The performance differences due to a change in retrieval@k remain mainly within a $\pm2\%$ range of the correctness evaluation metric (Figure~\ref{box:claude-eval}).
Increasing rerank@k with fixed retrieval@k results in a higher variation in the correctness score, where we measure variations from $1\%$ up to $25\%$.
The influence of the inverted context order averages to $1\%$ performance increase.

Table~\ref{tab:Gemma3-eval-500-basics} presents the results of the best-performing RAG settings for each approach, using identical settings of Pinecone@200, BGE@5 and inverted context order. IterDRAG uses Pinecone@200 and BGE@10 for initial retrieval, and Pinecone@200 with BGE@4 for up to 5 iterations.
With InstructRAG and IterDRAG perform comparably on Gemma-3-27B, we select both as possible approaches for LiveRAG.
During the live challenge day, we generated answers with both RAG solutions and evaluated the results using Gemma-3-27B, omitting the \textit{golden document} and \textit{golden answer}, supplemented by human evaluation.
Resulting in InstructRAG outperforms IterDRAG in terms of the evaluation metric. IterDRAG achieved an average correctness of $1.70$ and a faithfulness score of $0.73$, while InstructRAG achieved a correctness of $1.91$ ($+28.1\%$) and a faithfulness score of $0.93$ ($+27.4\%$). An additional manual comparison between these two approaches revealed an occasionally subjectively better question-answer alignment for InstructRAG.
Compared to our measurements, the organizer's LLM evaluation returned lower scores for our submitted InstructRAG-based approach: correctness of $1.13$ ($-40.9\%$) and faithfulness of $0.55$ ($-40.9\%$). This difference was likely influenced by using a more capable judge LLM and accessing the \textit{golden document} and \textit{golden answer}.

\begin{table}[t]
\centering
\caption{Gemma-3-27B evaluation on 500 DataMorgana-generated questions, equally distributed between single and multi-hop questions. We report the LiveRAG prompt (Fig.~\ref{box:claude-eval}) metrics [\%] for Correctness \{1, 2\} and Faithfulness \{0,1\}, discarding other Correctness \{-1, 0\} and Faithfulness \{-1\} values.}
\label{tab:Gemma3-eval-500-basics}
\setlength{\tabcolsep}{2pt}
\scalebox{0.93}{
\begin{tabular}{p{2cm} cc cc cc cc} 
\toprule
 & \multicolumn{4}{c}{Single-Hop}  & \multicolumn{4}{c}{Multi-Hop} \\
    \cmidrule(lr){2-5} \cmidrule(lr){6-9} 
 & \multicolumn{2}{c}{Correctness} & \multicolumn{2}{c}{Faithfulness} & \multicolumn{2}{c}{Correctness} & \multicolumn{2}{c}{Faithfulness} \\
RAG                 & \{ 1, & 2 \} & \{ 0, & 1 \} & \{ 1, & 2 \} & \{ 0, & 1 \} \\
\midrule
Simple Prompt       & 10.0 & 89.6 & 8.4 & 91.6 & 6.0 & 91.2 & 5.2 & 92.0 \\ 
TrustRAG            & 15.3 & 80.9 & 14.8 & 80.9 & 14.7 & 83.3 & 17.3 & 80.7 \\ 
Astute RAG          & 17.0 & 77.1 & 15.7 & 78.3 & 35.5 & 62.0 & 36.6 & 60.4 \\ 
InstructRAG         & 4.2 & \textbf{94.5} & 3.4 & \textbf{95.3} & 5.6 & 92.9 & 5.1 & 93.4 \\ 
IterDRAG            & 3.4 & \textbf{94.5} & 4.2 & 93.6 & 3.8 & \textbf{93.6} & 3.8 & \textbf{93.6} \\ 
\bottomrule
\end{tabular}}
\end{table}

\section{Conclusion and Future Work}\label{sec:conclusion}

For the LiveRAG Challenge, we investigated the performance parameters defined by the organizers for the evaluation, alongside variations of RAG components. We evaluated several RAG solutions, including AstuteRAG, IterDRAG, TrustRAG, InstructRAG, and simple prompting, using subsets of single- and multi-hop questions generated with DataMorgana. Our investigation included the performance of BGE and Rank-R1 rerankers, metrics for OpenSearch and Pinecone retrievers, and evaluations using Gemma-3-27B and Claude-3.5-Haiku as judge LLMs.
On the live challenge day, we ran our two best-performing RAG solutions in parallel, evaluating them with Claude-3.5-Haiku and manual assessment. Our final submitted solution uses InstructRAG with an inverted context order, employing Pinecone@200 with BGE@5 for context retrieval and reranking.
In the future, we plan to explore efficient and SotA RAG approaches to enhance performance on diverse QA datasets, starting with our DataMorgana-generated questions.



\bibliographystyle{ACM-Reference-Format}

\appendix
\section{Appendix}

\vspace{-0.5cm}
\begin{figure}[H]
    \caption{Pinecone excels OpenSearch in retrieving gold documents [\%], for multi-hop @k=20 and @k=50 for single-hop questions. Considering multi-hop questions, both documents used for QA generation must be retrieved.}
    \includegraphics[width=\columnwidth]{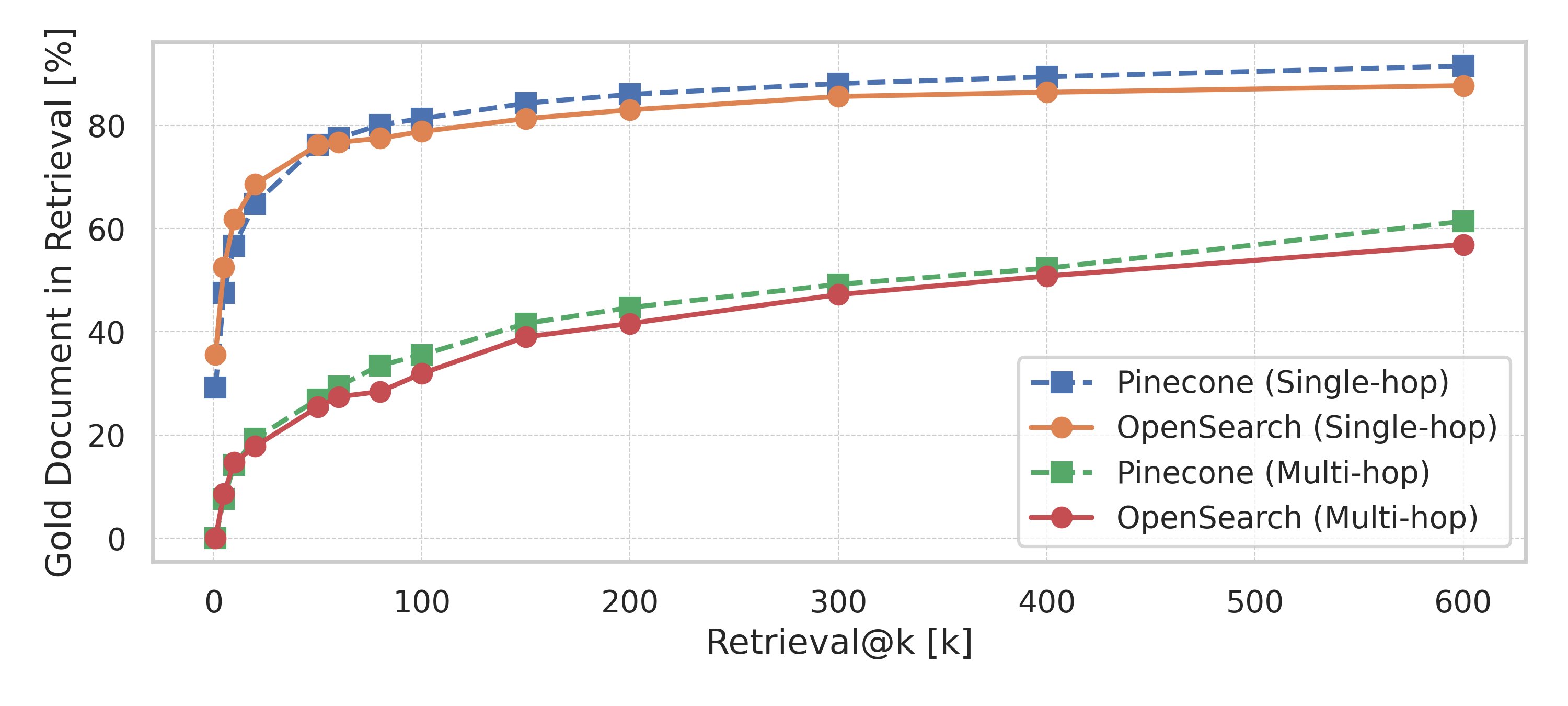}
    \label{fig:gold_docs_ret}
\end{figure}
\vspace{-2cm}
\begin{figure}[H]
    \caption{Gold documents [\%] in retrieval for Faclon3-10B context window of k=50 documents with Pinecone@k (76.2\%) compared to a selection of Pinecone@k$_1$ and BGE@k$_2$. Pinecone@80 with BGE@20 (76.7\%) reduces the context size while exceeding Pinecone@50 performance.}
    \includegraphics[width=\columnwidth]{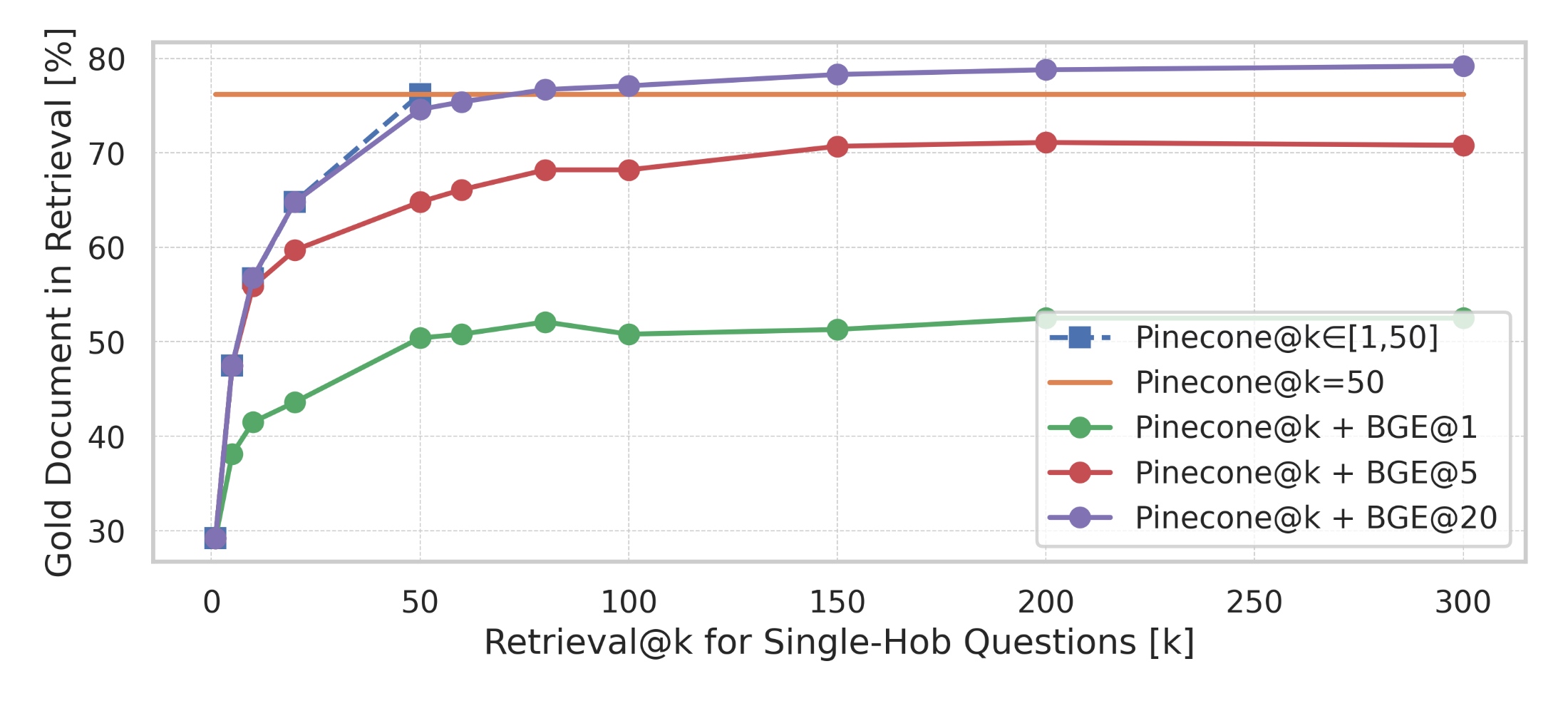}
    \label{fig:gold_docs_ret_2}
\end{figure}
\vspace{-2cm}
\begin{figure}[H]    
    \caption{OpenSearch and Pinecone retrievers, and BGE reranker scale linearly in runtime [s] for $k\in[1, 600]$ for our generated DataMorgana QA dataset.}
    \includegraphics[width=\columnwidth]{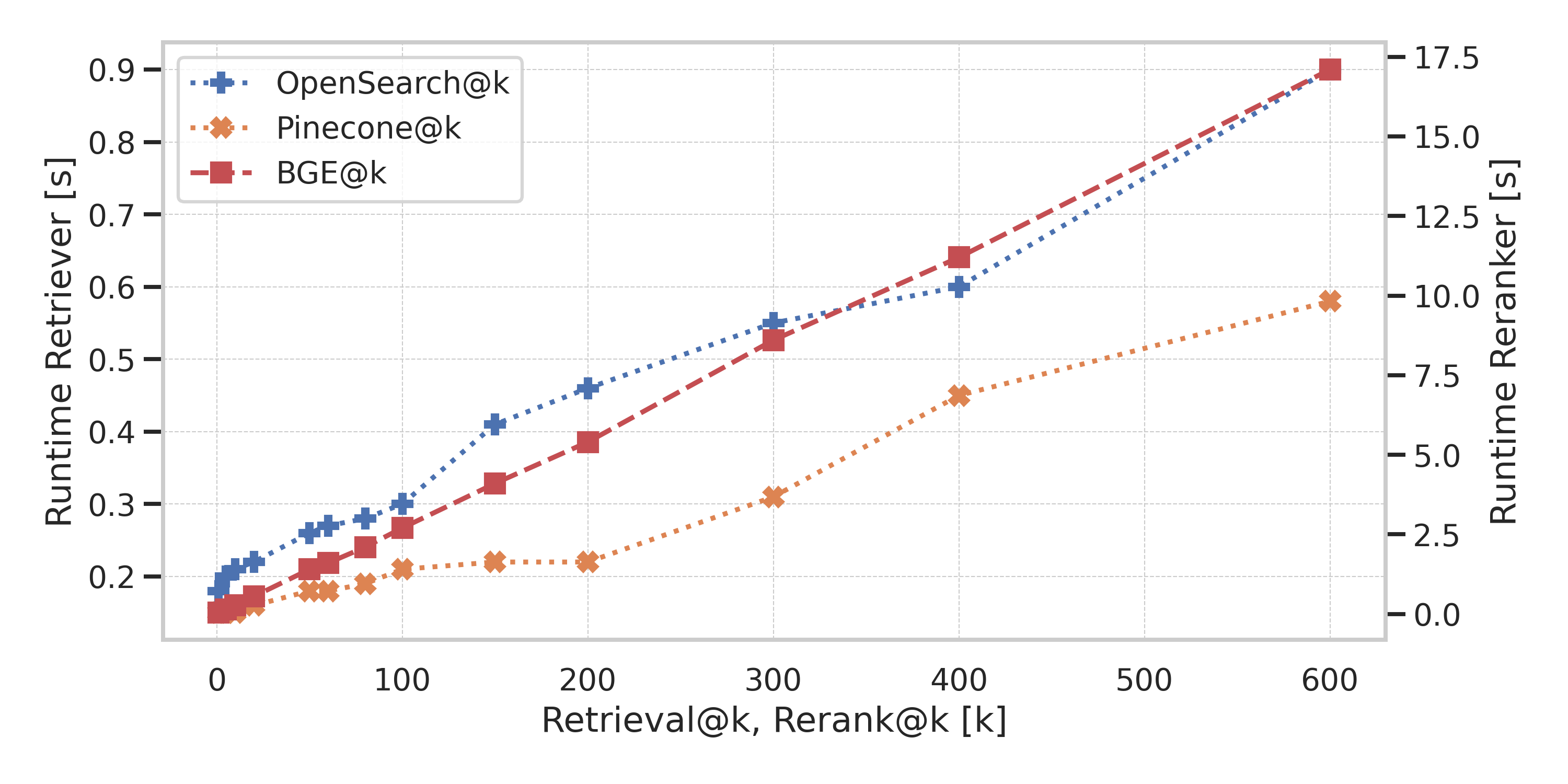}
    \label{fig:runtimes}
\end{figure}

\begin{minipage}{\columnwidth}
\captionof{figure}{Evaluation prompt based on LiveRAG evaluation guidelines used for Gemma-3-27B and Claude-3.5-Haiku.}
\label{box:claude-eval}
\begin{mybox}
\small
\textbf{System Prompt}\\
You are an expert evaluator assessing the quality of an answer to a given question based on retrieved passages.
Your evaluation must measure `correctness` and `faithfulness` according to predefined criteria.
\\\\
\textbf{Query Prompt}\\
Please evaluate the **generated answer** based on these specific metrics:

1. Correctness. Combines elements of:\\
\phantom{A}- **coverage**: portion of vital information, in the ground truth \\\phantom{AAA}answer which is covered by the generated answer.\\
\phantom{A}- **relevance**: portion of the generated response which is directly\\\phantom{AAA}addressing the question, regardless its factual correctness.\\
Graded on a continuous scale with the following representative points:\\
\phantom{A}- **2:** Correct and relevant (no irrelevant information)\\
\phantom{A}- **1:** Correct but contains irrelevant information\\
\phantom{A}- **0:** No answer provided (abstention)\\
\phantom{A}- **-1:** Incorrect answer\\

2. Faithfulness. Assesses whether the response is **grounded in the retrieved passages**.
Graded on a continuous scale with the following representative points:\\
\phantom{A}- **1:** Full support. All answer parts are grounded\\
\phantom{A}- **0:** Partial support. Not all answer parts are grounded\\
\phantom{A}- **-1:** No support. All answer parts are not grounded\\

The question that was asked:\\
\{question\}\\

Ground truth answer:\\
\{gold\_answer\}\\

Ground truth passage from a document:\\
\{gold\_document\}\\

Generated answer (to be evaluated):\\
\{generated\_answer\}\\

Retrieved passages:\\
\{retrieved\_passages\}\\

Return ONLY a JSON object with your evaluation scores. Do not repeat the question, answer or any other text. The JSON object should contain the following\\
\{\{"correctness": integer, "faithfulness": integer\}\}
\end{mybox}
\end{minipage}

\begin{minipage}{\columnwidth}
\captionof{figure}{Subset of question and user categories we used to generate QA-pairs with DataMorgana.}
\label{box:datamorgana}
\begin{mybox}
\small


\textbf{Question: quantitative-nature}\\
- Question that doesn't involve numbers, statistics, or mathematical reasoning. \\
- Question involving simple numerical facts or basic arithmetic (dates, measurements, counts). \\
- Question requiring statistical analysis, probability assessment, or mathematical reasoning beyond basic arithmetic.\\

\textbf{Question: verification-difficulty}\\
- Facts that can be quickly checked against reliable, accessible sources.\\
- Information requiring specialized sources or moderate effort to verify accurately. \\
- Information that's challenging to verify due to conflicting sources, limited documentation, or inherent ambiguity. \\

\textbf{User: decision-making-style}\\
- User who makes decisions based on logical analysis of facts and evidence.\\
- User who relies heavily on intuition and gut feelings when making decisions.\\
- User who prioritizes gathering multiple perspectives before making decisions.\\
- User who primarily draws on personal or historical experience to make decisions. \\

\textbf{User: trust-orientation}\\
- User who questions information extensively and requires strong evidence before accepting claims.\\
- User who verifies important information but doesn't question everything.\\
- User who generally trusts information from seemingly authoritative sources.\\
- User who rarely questions the validity of information they encounter. \\

\textbf{User: behavior-patterns}\\
- A spy usually hides his true intentions by omitting information, giving misleading details or communicating in encrypted form.\\
- A person who believes that this world is secretly controlled by reptilian aliens, often depicted as shape-shifting lizard-like beings. This belief is part of a conspiracy theory that suggests these beings manipulate human affairs and are involved in various global events. \\
- A member of a diverse group of reptiles that first appeared during the Triassic period, over 230 million years ago, and dominated the Earth for over 160 million years. \\
- Bishop of Rome and the spiritual leader of the worldwide Roman Catholic Church, he always refers to a god and is a religious figure. Cites biblical texts and religious doctrines in discussions, often emphasizing the importance of faith, morality, and the teachings of Jesus Christ. \\
- A person who claims to have traveled through time, either from the future or the past. They might offer vague predictions or anachronistic knowledge, often with a sense of urgency or a warning about future events. \\
\end{mybox}
\end{minipage}

\end{document}